\documentclass[twocolumn]{webofc}

\usepackage[varg]{txfonts}   
\usepackage{hyperref}
\usepackage{url}
\hypersetup{colorlinks=true,citecolor=blue,urlcolor=blue,linkcolor=blue}
\newcommand{\be}{\begin{eqnarray}}
\newcommand{\ee}{\end{eqnarray}}
\newcommand{\ba}{\begin{array}}
\newcommand{\ea}{\end{array}}
\newcommand{\bi}{\begin{itemize}}
\newcommand{\ei}{\end{itemize}}
\newcommand{\nn}{\nonumber}

\begin{document}
\title{
Improved phenomenology of $\pi N$ transition distribution amplitudes
}

\author{\firstname{Bernard} \lastname{Pire}\inst{1}\fnsep\thanks{\email{bernard.pire@polytechnique.edu}}
\and
        \firstname{Kirill} \lastname{Semenov-Tian-Shansky}\inst{2,3}\fnsep\thanks{\email{ksemenov@knu.ac.kr}} \and
        \firstname{Pawe{\l}} \lastname{Sznajder}\inst{4}\fnsep\thanks{\email{pawel.sznajder@ncbj.gov.pl}}
        \and
        \firstname{Lech} \lastname{Szymanowski}\inst{4}\fnsep\thanks{\email{lech.szymanowski@ncbj.gov.pl}}
\institute{Centre de Physique Th\'eorique, CNRS, \'Ecole Polytechnique, I.P. Paris, 91128 Palaiseau, France
\and
           Kyungpook National University, Daegu 41566, Korea
\and
        NRC ``Kurchatov Institute'' - PNPI, Gatchina 188300, Russia
\and
          National Centre for Nuclear Research, NCBJ, 02-093 Warsaw, Poland
          }
}
\abstract{To study cross sections and polarization asymmetries for the processes $e p \to e n \pi^+$ and $e p \to e p \pi^0$ in the backward region, we develop a flexible phenomenological  model for nucleon-to-pion transition distribution amplitudes ($\pi N$ TDAs), which are used in the QCD collinear factorization description of the scattering amplitudes. 
Our model is based on the two-component factorized Ansatz for the corresponding spectral densities, quadruple distribution. 
It takes into account the constraints for $\pi N$ TDAs arising from
the threshold pion production theorem and also includes a forward limit contribution that can be fitted to experimental data. 
We examine the sensitivity of observable predictions to various modelling assumptions.
}
\maketitle
\section{Introduction}
\label{intro}

Hard exclusive electroproduction reactions in the near-backward kinematics open a new window on hadronic structure, 
providing access to nucleon-to-meson (and nucleon-to-photon) transition distribution amplitudes (TDAs).
TDAs are defined through meson-nucleon matrix elements of the three-quark light-cone operator. 
They extend the concepts of both generalized parton distributions (GPDs) and nucleon distribution 
amplitudes (DAs), encoding correlations among the three valence quarks during a nucleon-to-meson transition.
This offers complementary tools for spatial and momentum-space tomography of hadrons. See Ref.~\cite{Pire:2021hbl} for a comprehensive review.

\begin{figure}[!ht]
\begin{center}
\includegraphics[width=0.25\textwidth]{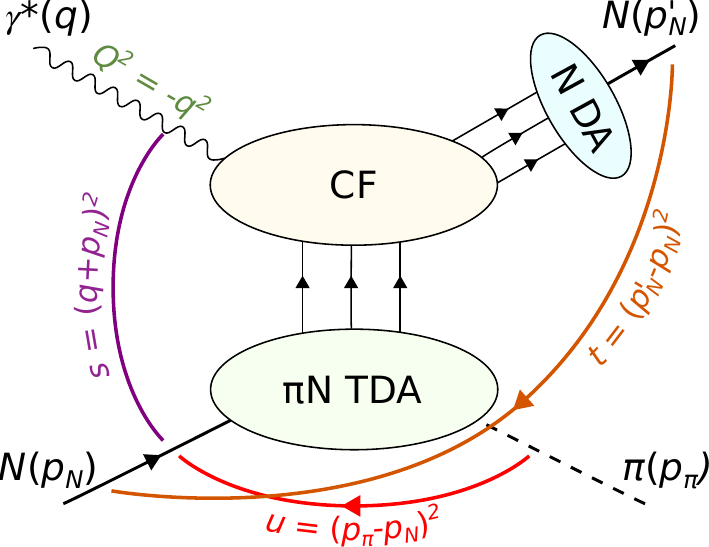}
\end{center}
\caption{Kinematical quantities and the collinear factorization mechanism for $\gamma^* N \to \pi N$ in the near-backward kinematical regime (large $Q^2$, $s$;
fixed $x_B$; $|u| \sim 0$). The lower blob, denoted $\pi N$ TDA, depicts the nucleon-to-pion  transition distribution amplitude; the $N$ DA blob depicts the
nucleon distribution amplitude; CF denotes the hard subprocess amplitude (coefficient function). Figure is taken from Ref.~\cite{Pire:2025wbf}.}
\label{Fig_Factorization}
\end{figure}

The collinear factorization mechanism for near-backward pion production off a nucleon is schematically presented in Fig.~\ref{Fig_Factorization}. 
The scattering amplitude is represented as a convolution of a perturbatively calculable hard part, referred to as the coefficient function (CF),
with nucleon DAs and the nucleon-to-pion ($\pi N$) TDAs. Although a rigorous proof of factorization for this process 
has not yet been established, the techniques 
\cite{Chen:2024fhj,Huang:2024ugd}
developed for the next-to-leading-order (NLO) analysis of the nucleon form factor 
may offer useful insights into this problem.

Backward hard exclusive reactions have  been experimentally challenged in several experiments at JLab. The pioneering study 
\cite{CLAS:2017rgp}
revealed, for the first time,  the manifestation of a peak in the backward scattering cross section for pion electroproduction. Furthermore,  the analysis 
of backward $\omega$-meson production demonstrated the dominance of the transverse cross section $\sigma_T$, providing  strong 
support for the relevance of the collinear factorization mechanism involving TDAs. Additional arguments in favour of the TDA framework come from the extraction of
the beam spin asymmetry (BSA) for hard exclusive $\pi^+$ production in a wide range of kinematics in Ref.~\cite{CLAS:2020yqf}.

An overview of proposed experiments to study backward hard exclusive reactions is provided in Ref.~\cite{Gayoso:2021rzj}. 
Specifically, this includes  a dedicated backward exclusive $\pi^0$ production experiment at JLab Hall C 
\cite{Li:2020nsk}. This highlights the need for a flexible phenomenological parametrization of 
$\pi N$ TDAs, enabling a reliable extraction of these quantities from experimental observables.  
We present recent advances~\cite{Pire:2025wbf} in the phenomenological modelling of $\pi N$ TDAs, formulated in a flexible yet theoretically constrained manner 
through the spectral representation in terms of quadruple distributions \cite{Pire:2010if}.

\section{Constructing and constraining models of TDAs}
\label{sec-1}

The spectral representation of a generic nucleon-to-meson TDA $H$ implements the
non-trivial interplay of the TDA dependence on the momentum fraction variables $x_{i}$, $i=1,\,2,\,3$,
and the skewness parameter $\xi$, defined with respect to the $u$-channel longitudinal momentum transfer:
\be
&&
\hspace{-3.2em} H \left(x_1, x_2, x_3, \xi \right) 
=   \left[\prod_{i=1}^3 \int_{\Omega_i} d \beta_i d \alpha_i\right]   \nn \\ && \hspace{-3.2em}\times \, \delta\left(x_1-\xi-\beta_1-\alpha_1 \xi\right)  
     \delta\left(x_2-\xi-\beta_2-\alpha_2 \xi\right) \nn \\ && 
\hspace{-3.2em}    
 \times \,     \delta\left( \sum_{i=1}^3 \beta_i
\right) \delta\left( \sum_{i=1}^3 \alpha_i
+1\right)   
 F\left(\beta_1, \beta_2, \beta_3, \alpha_1, \alpha_2, \alpha_3 \right),
 \label{Spectral_TDA}
\ee
where $\Omega_i=\left\{\left|\beta_i\right| \leq 1 \cap\left|\alpha_i\right| \leq 1-\left|\beta_i\right|\right\}$ 
denote three copies of the standard square spectral region. The spectral density $F$  is a function of six spectral variables that are subject to two constraints imposed by the two $\delta$-functions in the last line of (\ref{Spectral_TDA}); thus, effectively, it is a quadruple distribution.
The spectral representation (\ref{Spectral_TDA}) ensures the polynomiality property of TDAs \cite{Pire:2011xv},
and accounts for the support properties of TDAs. Particularly, in the ``forward limit,'' $\xi=0$,
the support is reduced to the regular hexagon, which is the usual support domain of the forward-type 3-body structure functions, see {e.g.}
Ref.~\cite{Rodini:2022wki}, while in the limit $\xi=1$ one recovers the standard equilateral triangle support of baryon DAs.

\begin{figure}[!ht]
\begin{center}
\includegraphics[width=0.48\textwidth]{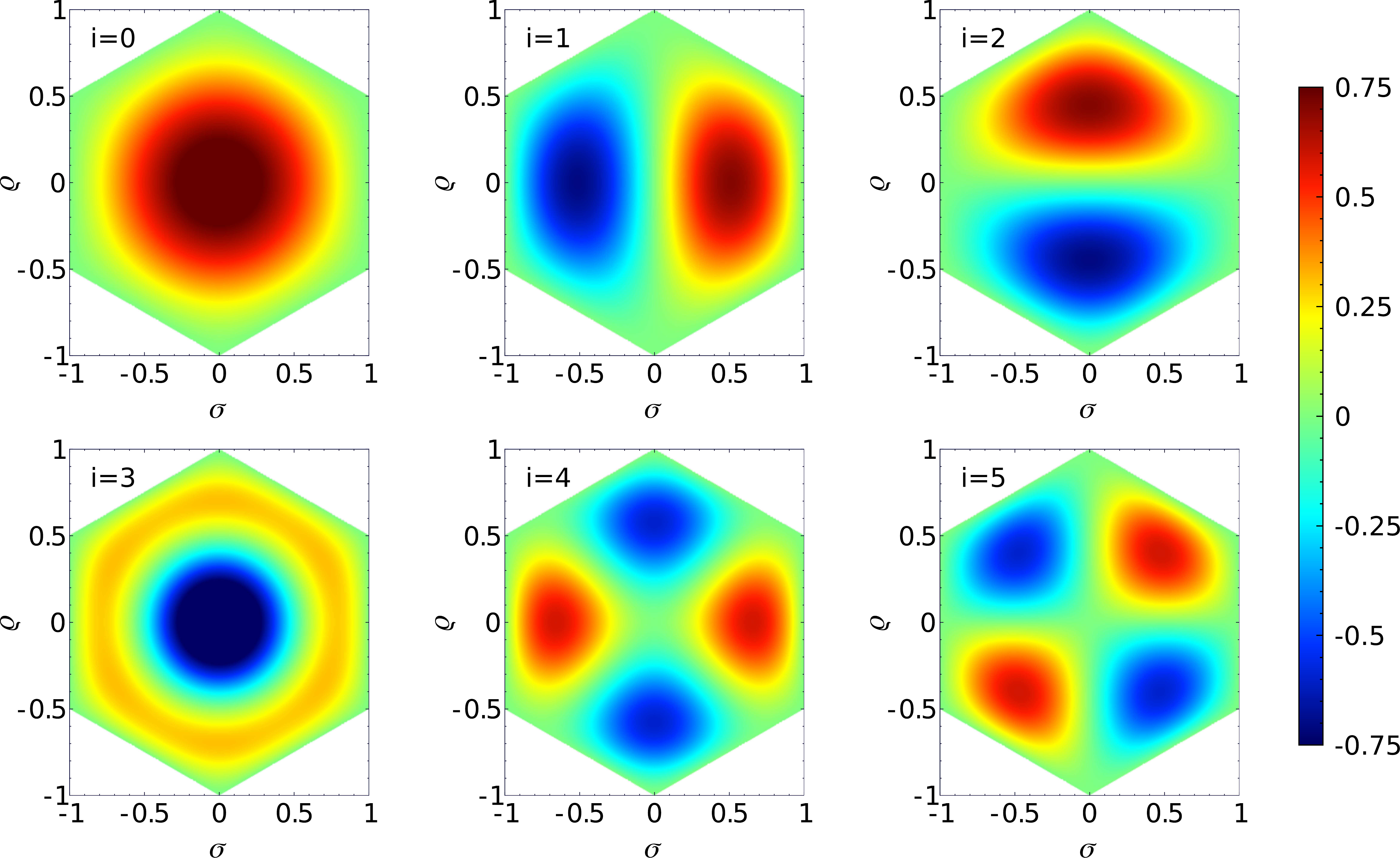}
\end{center}
\caption{Orthogonal polynomials on the hexagon, multiplied by the weight function (\ref{eq:def_weight}) with $d=1$, $W(\sigma, \rho)\,p_{i}(\sigma,\rho)$, for $i=0,\ldots,5$. Figure is taken from Ref.~\cite{Pire:2025wbf}.}
\label{fig:fig_Hex}
\end{figure}

The factorized Ansatz for double distributions (DDs) \cite{Musatov:1999xp},
presenting DDs as a product of a forward PDF $q(\beta)$ and a universal profile function $h(\beta,\alpha)$,
has been instrumental in the GPD phenomenology. We provide a generalization of this idea for the case of
quadruple distributions.   
In order to reduce the intrinsic redundancy in the kinematical description of a 3-quark system, 
it is convenient to switch to the so-called 
quark-diquark coordinates, $w \equiv x_3-\xi$; $v \equiv (x_1-x_2)/2$, and employ the associated combinations of 
the spectral variables 
$\sigma \equiv \beta_3$; $ \rho  \equiv (\beta_1-\beta_2)/2$; $\omega \equiv \alpha_3$; $  \nu \equiv (\alpha_1-\alpha_2)/2$.
This puts the quadruple spectral density $ F (\sigma, \rho, \omega, \nu)  \equiv F\left(\rho-\frac{\sigma}{2},-\rho-\frac{\sigma}{2}, \sigma, \nu-\frac{1+\omega}{2},-\nu-\frac{1+\omega}{2}, \omega\right)
$ 
in the form suitable to construct the factorized Ansatz:
\be
F (\sigma, \rho, \omega, \nu) = f(\sigma, \rho) \times h(\sigma, \rho, \omega, \nu). 
\label{Fact_Ansatz}
\ee
Here $f(\sigma, \rho)$ corresponds to the forward limit of the TDA defined on the hexagon
$|\sigma|\le 1; \, |\rho| \le 1-|\sigma|$,
and the normalized profile function is chosen as 
\be
& h(\sigma,\,\rho,\, \omega,\, \nu) =
 \frac{\Gamma(3b+3)}{2^{5b+2} \Gamma(1+b)^3} \nonumber \\ &
\times \frac{\left(1+2 \nu -\omega -2 \left|\rho -\frac{\sigma }{2}\right|\right)^b \left(1-2 \nu -\omega -2 \left|\rho +\frac{\sigma}{2}\right| \right)^b (1 -|\sigma |+\omega)^b}
{\left(1-\frac{1}{2} \left(\left|\rho -\frac{\sigma }{2}\right|+\left|\rho+\frac{\sigma }{2}\right|+|\sigma |\right) \right)^{3b+2}} \,,
\label{eq:profile_function}
\ee
where the parameter $b$ controls the strength of the skewness dependence.
However, contrary to the GPD case,  the forward function $f(\sigma,\rho )$ is not related to any directly measured quantity. 
To construct a flexible parametrization of the forward function, we use a basis $p_i(\sigma,\rho)$ of orthogonal polynomials  on the hexagon, shown in Fig.~\ref{fig:fig_Hex}, 
which has been devised in the context of hexagonal optical elements \cite{Hexagon}:
\begin{align}
f(\sigma, \rho)= W(\sigma, \rho)
\sum_{i=0}^{n(n+3)/2}
c_{i}\, p_{i} (\sigma, \rho)
\,.
\label{eq:forward_limit_ansatz}
\end{align}
The number of free
parameters is controlled by $n$, the maximum
total power of polynomials taken into account.
To ensure good convergency properties at the borders of the domain of definition, we choose the weight function $W(\sigma,\rho)$, that defines the orthogonal basis, as
\begin{equation*}
W(\sigma, \rho) = N \left(1-\sigma ^2\right)^d \left((\rho -1)^2-\frac{\sigma^2}{4}\right)^d \left((\rho +1)^2-\frac{\sigma ^2}{4}  \right)^d,
\label{eq:def_weight}
\end{equation*}
depending on a parameter $d$. 
Thus, employing (\ref{eq:profile_function}) and (\ref{eq:forward_limit_ansatz}), we obtain a flexible Ansatz for generic TDAs with input at \mbox{$\xi=0$}.

For the case of $\pi N$ TDAs, we also would like to implement the constraints arising from the threshold pion production theorem.
For that purpose, we generalize the framework of the two-component model for DDs proposed in Ref.~\cite{Teryaev:2001qm} and
construct the two-component model for quadruple distributions: 
\be
F(\ldots)= F^{(0)}(\ldots)+ (1-\xi)F^{(1)}(\ldots).
\ee
The spectral density $F^{(0)}$  accounts for the $\xi=1$ limit of TDAs that is fixed through the threshold pion production theorem
\cite{Pobylitsa:2001cz,Braun:2006td} 
in terms of combinations of nucleon DAs. It is constructed along the lines of Ref.~\cite{Lansberg:2011aa}. For the second term, that 
vanishes in the  $\xi=1$ limit,  we adopt the flexible parametrization based on the factorized Ansatz 
(\ref{Fact_Ansatz}), (\ref{eq:profile_function}), (\ref{eq:forward_limit_ansatz}). The latter is further improved implementing the factorized form of
the $u$-dependence, described by a
modified dipole formula,  $G(u)=\left(1-\frac{u-m_N^2}{m_D^2}\right)^{-2}$.
The free parameters in 
(\ref{eq:forward_limit_ansatz}), 
$c_i$, can be constrained by fitting to experimental
data. Since only very limited data is available at this time, we
set $n=1$, and only consider $c_0$, $c_1$ and $c_2$ choosing the specific ratio $c_0: c_1: c_2=1: 1: 1$.
The value of $c_0$ is then determined to replicate the data for the unseparated  near-backward 
$e p \rightarrow e^{\prime} n \pi^{+}$ 
collected by CLAS
\cite{CLAS:2017rgp}.
We employ the so-called ``unseparated'' cross section
$\sigma_U=\sigma_T+ \varepsilon \sigma_T$, with $\varepsilon$ denoting the 
the polarization parameter, 
assuming the relevance of the collinear factorization mechanism  
$\sigma_T \gg \sigma_L$. 
Thus, once the model parameters are fixed as described above, we can make quantitative predictions for the cross sections and polarization observables 
of the near-backward reactions 
$e p \rightarrow e^{\prime} n \pi^{+}$ 
and
$e p \rightarrow e^{\prime} p \pi^{0}$
for the kinematical conditions typical at JLab. 

%

\begin{figure*}[h]
\centering
\includegraphics[width=0.25\textwidth]{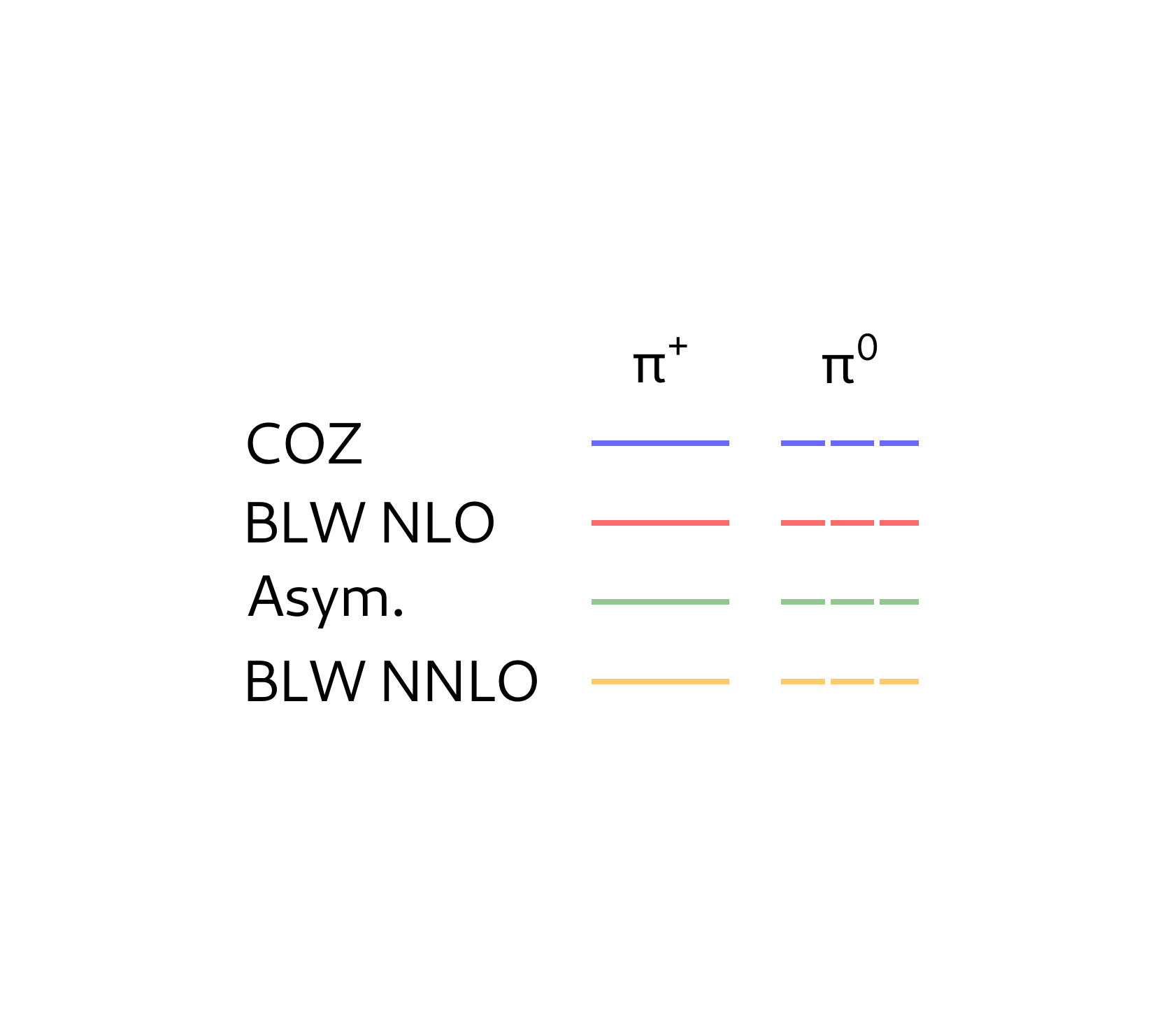}
\includegraphics[width=0.25\textwidth]{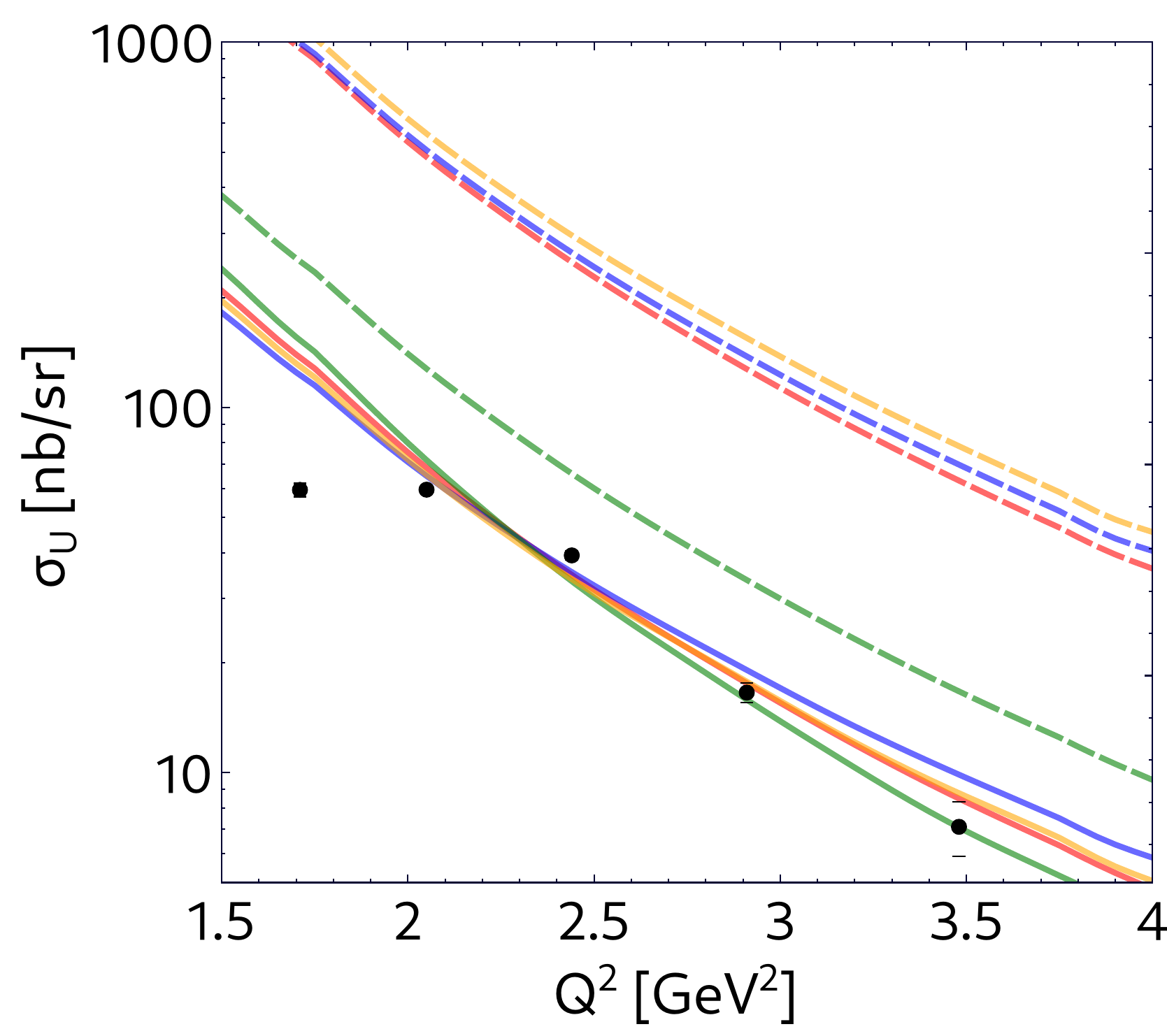}
\includegraphics[width=0.25\textwidth]{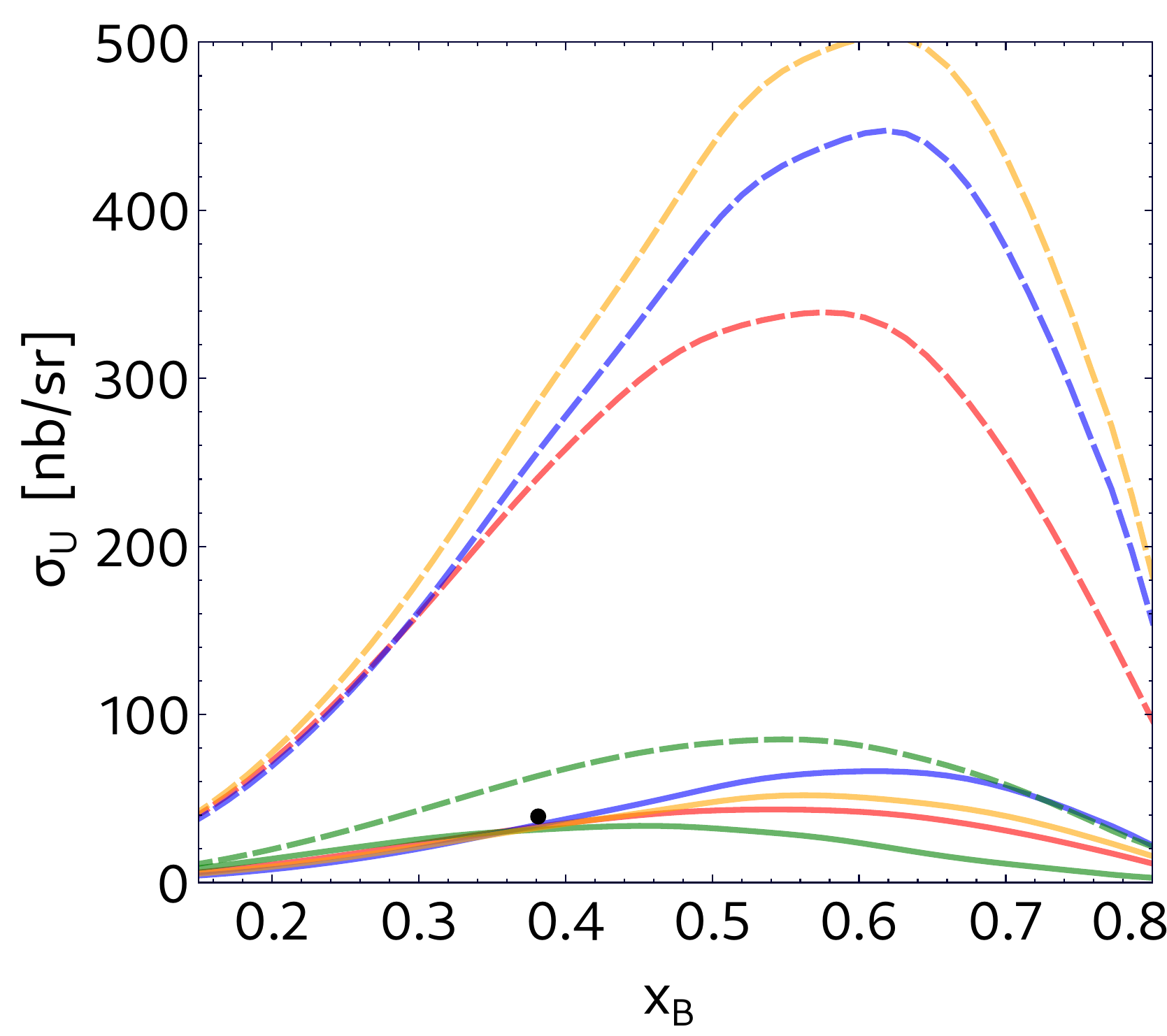}
\\
\includegraphics[width=0.25\textwidth]{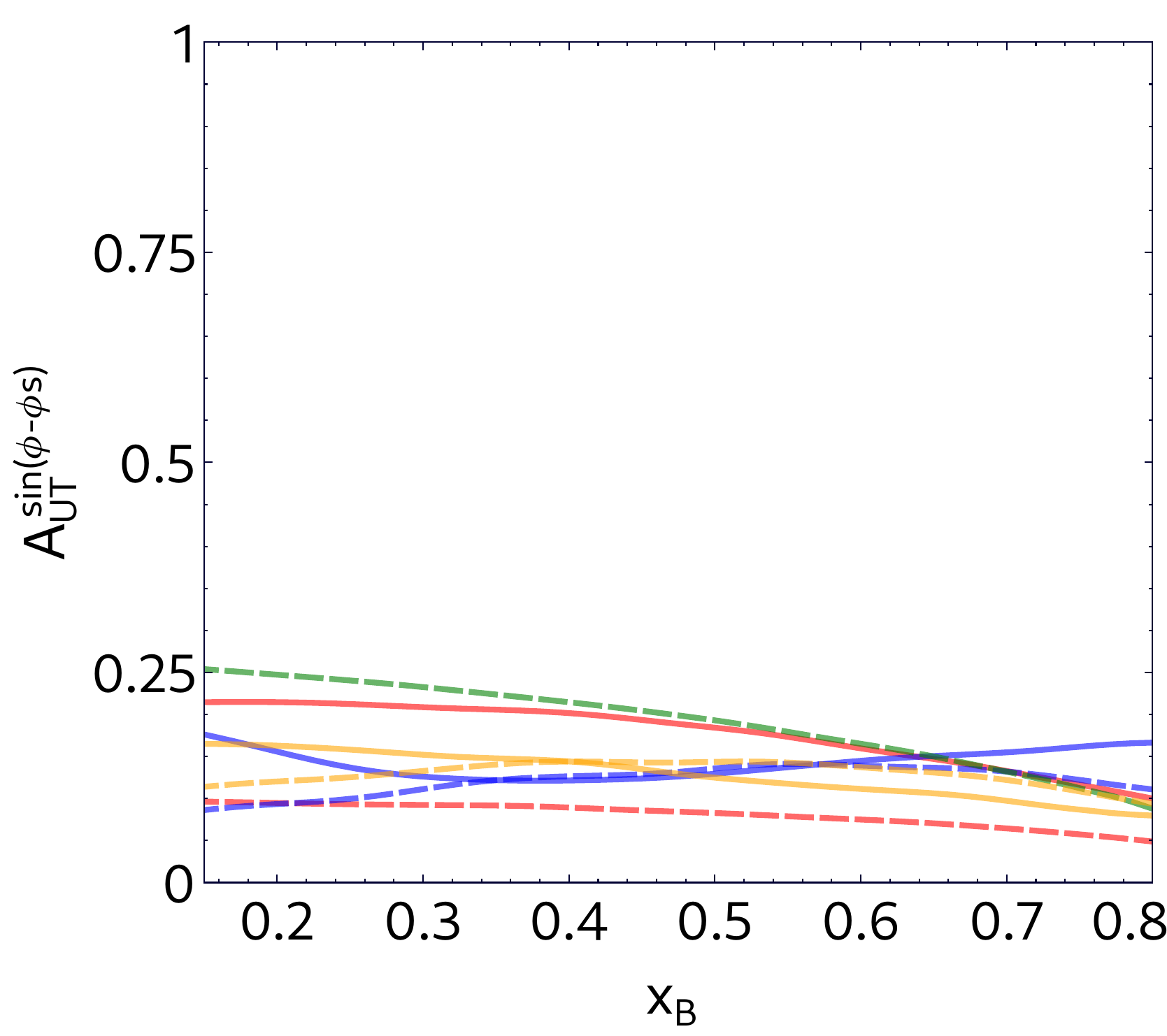}
\includegraphics[width=0.25\textwidth]{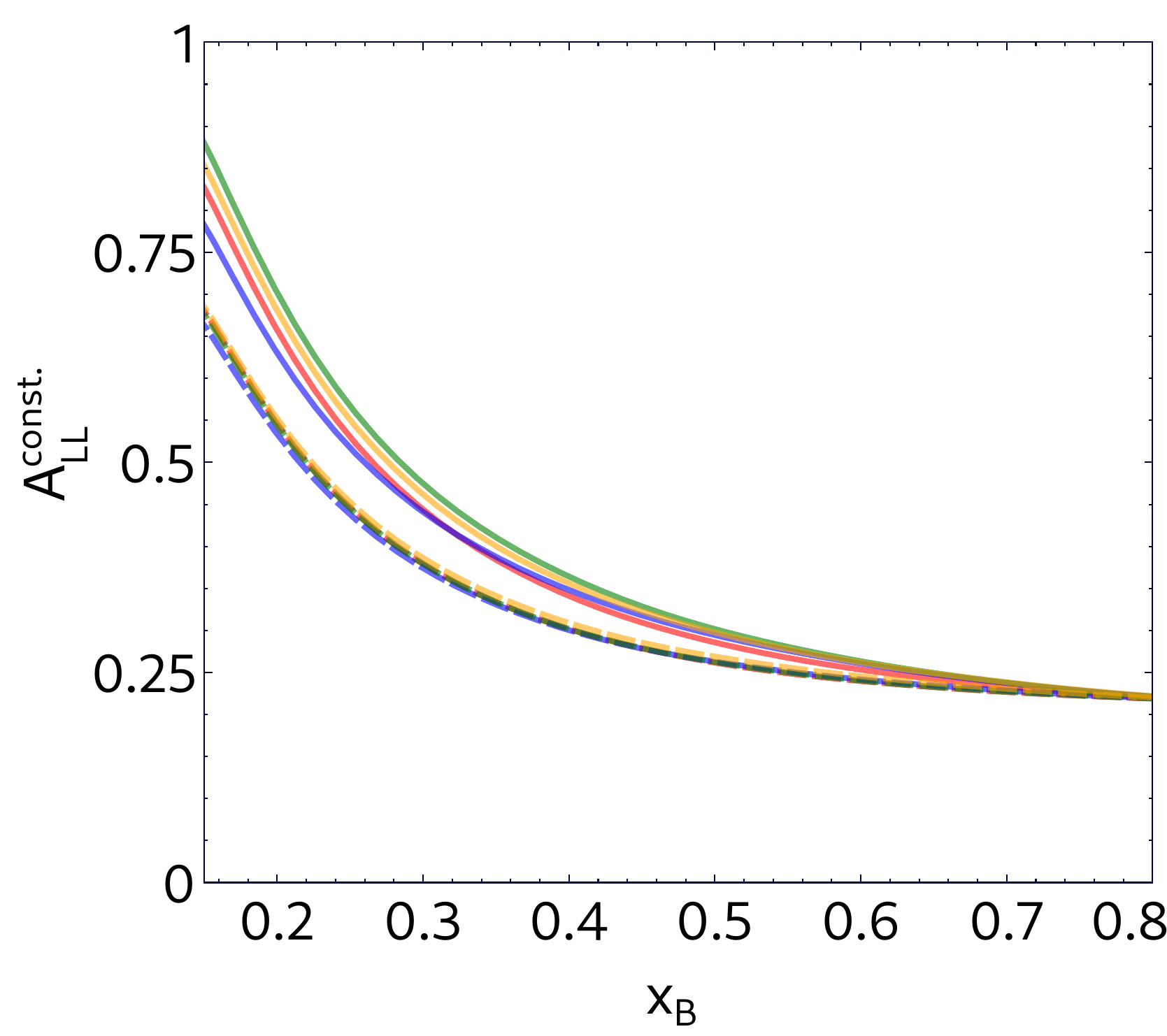}
\includegraphics[width=0.25\textwidth]{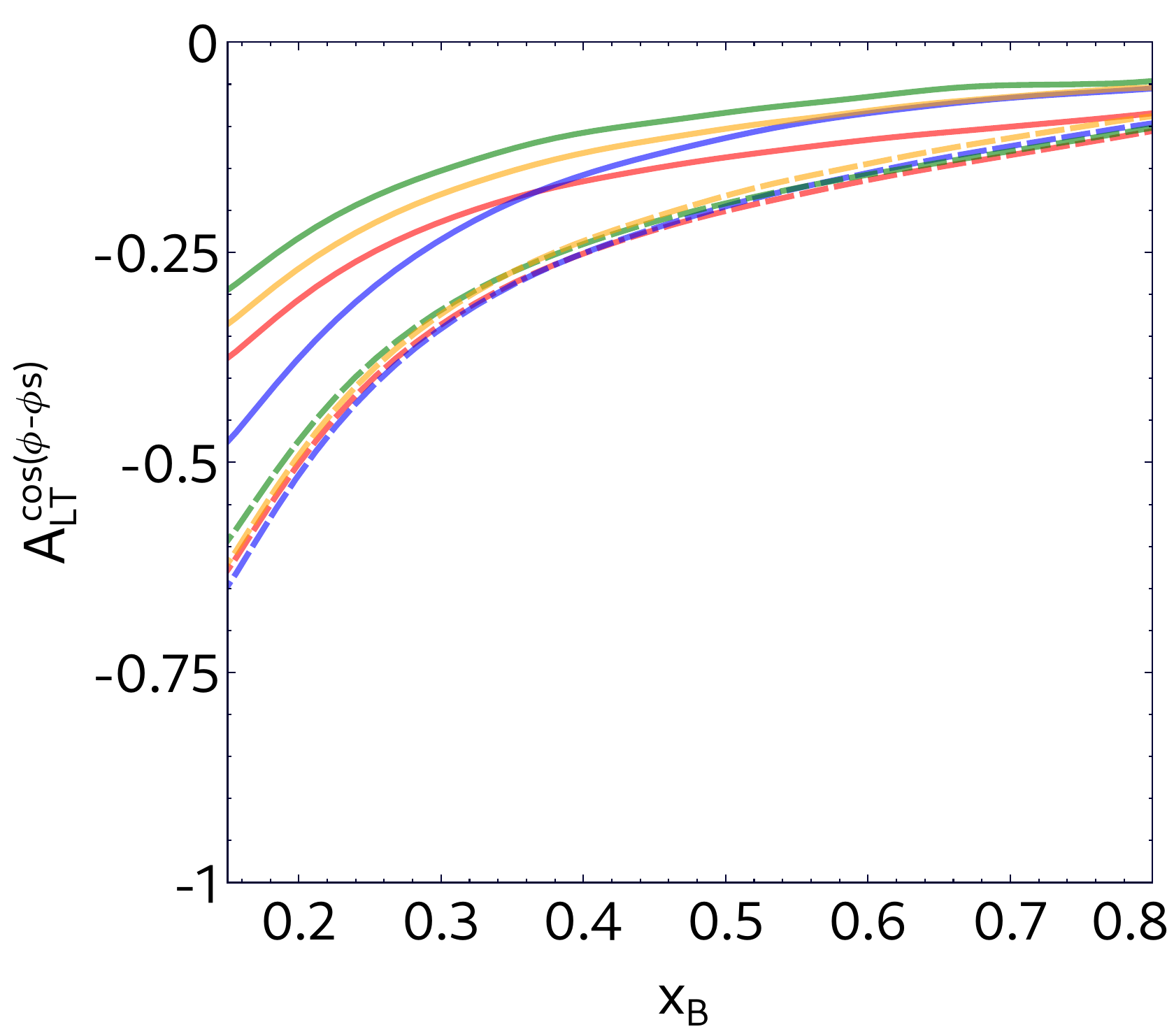}
\caption{Sensitivity to the choice of DA parametrization. Selected observables for the exclusive production of (solid lines) $\pi^+$ and (dashed lines) $\pi^0$ mesons in the backward kinematics are: cross sections as a function of $Q^2$ for $W = 2.2~\mathrm{GeV}$ and $-u = 0.5~\mathrm{GeV}^2$; cross sections and asymmetries
as a function of $x_B$ for $Q^2 = 2.44~\mathrm{GeV}^2$ and $-u = 0.5~\mathrm{GeV}^2$. The asymmetries are evaluated for the electron beam energy $E_e = 10.6~\mathrm{GeV}$. The CLAS data~\cite{CLAS:2017rgp} are represented by black markers. Figure is taken from Ref.~\cite{Pire:2025wbf}.}
\label{fig:scan_da}
\end{figure*}

\section{Cross sections and asymmetries}

The experimentally measured 5-fold cross section of the $eN \to e'N'\pi$ reaction 
can be put in one-to-one correspondence with the helicity amplitudes
$\mathcal{M}_{s_N ; s_N^{\prime}}^{\lambda_\gamma=m}$ of the hard subprocess 
$\gamma^* N \rightarrow \pi N^{\prime}$, see Ref.~\cite{Pire:2025wbf} and also
Ref.~\cite{Diehl:2005pc}.

Within the collinear factorization framework involving
$\pi N$ TDAs, with the reaction mechanism depicted in Fig.~\ref{Fig_Factorization},
only a few photoabsorption cross sections and interference
terms obtain contributions to the leading
twist-3 accuracy.  Nevertheless,  one
can construct several polarization observables that receive leading twist contributions, providing a test for the validity of the proposed reaction mechanism.
These include:
\begin{itemize}
\item The transverse target spin single asymmetry
\be
A_{U T}=\frac{d \sigma^{\Uparrow}-d \sigma^{\Downarrow}}{d \sigma^{\Uparrow}+d \sigma^{\Downarrow}},
\ee
sensitive to the $\sin(\phi-\phi_S)$-modulated term, where $\phi$ is the angle between the leptonic and hadronic 
reaction planes and   $\phi_S$ is the azimuthal angle of the transverse component
of the target nucleon spin;

\item The longitudinal-beam-longitudinal target
double spin asymmetry,
\be
\!\!\! \!\!\! A_{LL}
= \frac{\left(d\sigma^{\rightarrow \Rightarrow}-d\sigma^{\leftarrow \Rightarrow}\right)-\left(d\sigma^{\rightarrow \Leftarrow}-d\sigma^{\leftarrow \Leftarrow}\right)}{
 d\sigma^{\rightarrow \Rightarrow}+d\sigma^{\rightarrow \Leftarrow}+d\sigma^{\leftarrow \Leftarrow}+d\sigma^{\leftarrow \Rightarrow}
}\,,
\label{Def_ALLDSA}
\ee
originating from the $\phi$-independent interference term with longitudinal-beam-longitudinal target polarization;

\item And the longitudinal-beam-transverse-target double
spin asymmetry,
\be
A_{LT} =
\frac{\left(d\sigma^{\rightarrow \Uparrow}-d\sigma^{\rightarrow \Downarrow}\right)+\left(d\sigma^{\leftarrow \Downarrow}-d\sigma^{\leftarrow \Uparrow}\right)}{d\sigma^{\rightarrow \Uparrow}+d\sigma^{\leftarrow \Downarrow}+d\sigma^{\rightarrow \Downarrow}+d\sigma^{\leftarrow \Uparrow}}\,,
 \label{Def_ALTDSA}
\ee
that arises from the $\cos \left(\phi-\phi_S\right)$-modulated term with longitudinal-beam-transverse-target polarization. 
\end{itemize}
These asymmetries are non vanishing at leading twist. Checking that they do not diminish at higher $Q^2$ values, contrarily to the case for forward meson electroproduction,
is an important test of the TDA picture. We demonstrated that these asymmetries are not parametrically small, under reasonable model assumptions for $\pi N$ TDAs.

\begin{figure}[!ht]
\begin{center}
\includegraphics[width=0.49\textwidth]{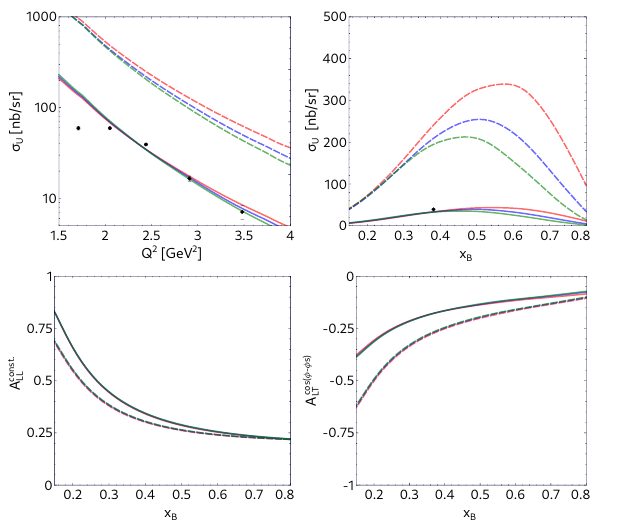}
\end{center}
\caption{Various model predictions ($d=1,2,3$) for the $Q^2$ and $x_B-$dependences of the cross section (upper row) and for the two leading-twist double spin asymmetries (lower row) for both $\pi^+$ (solid curves) and $\pi^0$ (dashed curves) electroproduction. See Captions of Fig.~\ref{fig:scan_da} for other details. Figure is taken from Ref.~\cite{Pire:2025wbf}.}
\label{fig:results}
\end{figure}

We illustrate the application of our approach by showing in Fig.~\ref{fig:scan_da} the sensitivity of the cross section and of polarization 
observables to the choice of nucleon DA parametrization including the COZ \cite{Chernyak:1987nv}, BLW NLO \cite{Braun:2006hz} and the BLW NNLO \cite{Lenz:2009ar} as well as the asymptotic nucleon DA. Black dots show the CLAS data \cite{CLAS:2017rgp} for the  $\pi^+$  channel used to normalize our phenomenological model for $\pi N$ TDAs.
We show results for both $\pi^+$ (solid lines) and $\pi^0$ (dashed curves) electroproduction in the backward region.
Fig.~\ref{fig:results} presents the sensitivity of the unpolarized cross section and of the double polarization asymmetries 
(\ref{Def_ALLDSA}) and (\ref{Def_ALTDSA}) on the parameter $d$ of the weight function (\ref{eq:def_weight}).

We would like to emphasize that only the normalization component has been fitted, while the other features of the model remain subject to assumptions.
Therefore, our predictions should be viewed as outcomes of a data-driven model rather than the result of a fully robust fit.
This leads to two key conclusions: (i) the $\pi^+$ and $\pi^0$ channels offer complementary insights into TDAs; and (ii) measuring of the $\pi^0$
channel will be significant, as it will provide distinct information about TDAs.
Moreover, if the experimental setup allows, measurements of single- and double-polarization observables of the $\pi^0$
channel would offer stringent tests of the TDA framework.

The study is complemented by the implementation of the exclusive $\pi^0$ meson production in backward kinematics in the Monte Carlo generator EpIC~\cite{Aschenauer:2022aeb}. This extension supplements the existing implementation of the process in forward kinematics (sensitive to GPDs), enabling a unified study of both kinematic regimes. The implementation also provides experimentalists with a practical tool for evaluating acceptance corrections in future measurements and for developing new experimental proposals. Additional details, including a demonstration of event generation for JLab 12 GeV kinematics, are presented in Ref.~\cite{Pire:2025wbf}.

\section{Conclusion}
Our new approach to the modelling of TDAs opens  the way for a rich phenomenology of  the processes $ep \to en\pi^+$ and $ep \to ep\pi^0$. TDAs are well-defined objects within the QCD collinear factorization framework and are complementary to GPDs. They offer valuable insights into the transition of a hadron into different particles, encoding correlations between partons during these transitions. However, the experimental program related to TDAs remains largely exploratory and underdeveloped, which directly motivated our study. Notably, dedicated experiments \cite{Li:2020nsk, Huber:2022wns} are proposed or planned at JLab, and additional data could be gathered from various setups in electroproduction, as well as in photoproduction or with meson and antiproton beams~\cite{PANDA:2016scz,Pire:2022kwu}.

The work of K.~S. was supported by Basic Science
Research Program through the National Research
Foundation of Korea (NRF) funded by the Ministry of
Education RS-2023-00238703 and RS-2018-NR031074.
The work of P.~S. and L.~S. was supported by
the Grant No. 2024/53/B/ST2/00968 of the National
Science Centre. 



%
\bibliography{bibliography} 
%
\end{document}